\documentclass[aps,superscriptaddress]{revtex4}
\usepackage{epsfig}
\usepackage{fleqn}
\usepackage{amssymb,amsmath}

\begin{document}        

\title{Measurement of the $\mathbf{e^+e^-\to K^+K^-}$ cross section
in the energy range $\mathbf{\sqrt{s}=1.05-2.0}$ GeV}

\begin{abstract}
The $e^+e^-\to K^+K^-$ cross section is measured in the center-of-mass energy 
range $1.05-2.00$ GeV at the SND detector. The measurement is
based on data with an integrated luminosity of 35 pb$^{-1}$ collected at
the VEPP-2000 $e^+e^-$-collider. The obtained results are consistent with the 
previous most accurate data obtained in the BABAR experiment
and have a comparable accuracy.
\end{abstract}

\author{M.~N.~Achasov}
\affiliation{Budker Institute of Nuclear Physics, 
630090 Novosibirsk, Russia}
\affiliation{Novosibirsk State University,
630090 Novosibirsk, Russia}

\author{V.~M.~Aulchenko}
\affiliation{Budker Institute of Nuclear Physics, 
630090 Novosibirsk, Russia}
\affiliation{Novosibirsk State University,
630090 Novosibirsk, Russia}

\author{A.~Yu.~Barnyakov}
\affiliation{Budker Institute of Nuclear Physics, 
630090 Novosibirsk, Russia}
\affiliation{Novosibirsk State University,
630090 Novosibirsk, Russia}

\author{M.~Yu.~Barnyakov}
\affiliation{Budker Institute of Nuclear Physics, 
630090 Novosibirsk, Russia}
\affiliation{Novosibirsk State University,
630090 Novosibirsk, Russia}

\author{K.~I.~Beloborodov}
\email[e-mail: ]{K.I.Beloborodov@inp.nsk.su}
\affiliation{Budker Institute of Nuclear Physics, 630090 Novosibirsk, Russia}
\affiliation{Novosibirsk State University, 630090 Novosibirsk, Russia}

\author{A.~V.~Berdyugin}
\affiliation{Budker Institute of Nuclear Physics, 
630090 Novosibirsk, Russia}
\affiliation{Novosibirsk State University, 630090 Novosibirsk, Russia}

\author{D.~E.~Berkaev}
\affiliation{Budker Institute of Nuclear Physics, 
630090 Novosibirsk, Russia}
\affiliation{Novosibirsk State University,
630090 Novosibirsk, Russia}

\author{A.~G.~Bogdanchikov}
\affiliation{Budker Institute of Nuclear Physics, 
630090 Novosibirsk, Russia}

\author{A.~A.~Botov}
\affiliation{Budker Institute of Nuclear Physics, 
630090 Novosibirsk, Russia}

\author{A.~R.~Buzykaev}
\affiliation{Budker Institute of Nuclear Physics, 
630090 Novosibirsk, Russia}
\affiliation{Novosibirsk State University,
630090 Novosibirsk, Russia}

\author{T.~V.~Dimova}
\affiliation{Budker Institute of Nuclear Physics, 
630090 Novosibirsk, Russia}
\affiliation{Novosibirsk State University,
630090 Novosibirsk, Russia}

\author{V.~P.~Druzhinin}
\affiliation{Budker Institute of Nuclear Physics, 
630090 Novosibirsk, Russia}
\affiliation{Novosibirsk State University,
630090 Novosibirsk, Russia}

\author{V.~B.~Golubev}
\affiliation{Budker Institute of Nuclear Physics, 
630090 Novosibirsk, Russia}
\affiliation{Novosibirsk State University,
630090 Novosibirsk, Russia}

\author{L.~V.~Kardapoltsev}
\affiliation{Budker Institute of Nuclear Physics, 
630090 Novosibirsk, Russia}
\affiliation{Novosibirsk State University,
630090 Novosibirsk, Russia}

\author{A.~G.~Kharlamov}
\affiliation{Budker Institute of Nuclear Physics, 
630090 Novosibirsk, Russia}
\affiliation{Novosibirsk State University,
630090 Novosibirsk, Russia}

\author{S.~A.~Kononov}
\affiliation{Budker Institute of Nuclear Physics,
630090 Novosibirsk, Russia}
\affiliation{Novosibirsk State University,
630090 Novosibirsk, Russia}

\author{I.~A.~Koop}
\affiliation{Budker Institute of Nuclear Physics, 
630090 Novosibirsk, Russia}
\affiliation{Novosibirsk State University,
630090 Novosibirsk, Russia}
\affiliation{Novosibirsk State Technical University, 
630092 Novosibirsk, Russia}

\author{A.~A.~Korol}
\affiliation{Budker Institute of Nuclear Physics, 
630090 Novosibirsk, Russia}
\affiliation{Novosibirsk State University,
630090 Novosibirsk, Russia}

\author{S.~V.~Koshuba}
\affiliation{Budker Institute of Nuclear Physics, 
630090 Novosibirsk, Russia}

\author{D.~P.~Kovrizhin}
\affiliation{Budker Institute of Nuclear Physics, 
630090 Novosibirsk, Russia}
\affiliation{Novosibirsk State University,
630090 Novosibirsk, Russia}

\author{E.~A.~Kravchenko}
\affiliation{Budker Institute of Nuclear Physics, 
630090 Novosibirsk, Russia}
\affiliation{Novosibirsk State University,
630090 Novosibirsk, Russia}

\author{A.~S.~Kupich}
\affiliation{Budker Institute of Nuclear Physics, 
630090 Novosibirsk, Russia}

\author{A.~P.~Lysenko}
\affiliation{Budker Institute of Nuclear Physics, 
630090 Novosibirsk, Russia}

\author{K.~A.~Martin}
\affiliation{Budker Institute of Nuclear Physics, 
630090 Novosibirsk, Russia}

\author{A.~E.~Obrazovsky}
\affiliation{Budker Institute of Nuclear Physics, 
630090 Novosibirsk, Russia}

\author{A.~P.~Onuchin}
\affiliation{Budker Institute of Nuclear Physics,
630090 Novosibirsk, Russia}
\affiliation{Novosibirsk State University,
630090 Novosibirsk, Russia}
\affiliation{Novosibirsk State Technical University, 
630092 Novosibirsk, Russia}

\author{A.~V.~Otboyev}
\affiliation{Budker Institute of Nuclear Physics, 
630090 Novosibirsk, Russia}

\author{E.~V.~Pakhtusova}
\affiliation{Budker Institute of Nuclear Physics, 
630090 Novosibirsk, Russia}

\author{E.~A.~Perevedentsev}
\affiliation{Budker Institute of Nuclear Physics, 
630090 Novosibirsk, Russia}
\affiliation{Novosibirsk State University,
630090 Novosibirsk, Russia}

\author{Yu.~A.~Rogovsky}
\affiliation{Budker Institute of Nuclear Physics, 
630090 Novosibirsk, Russia}
\affiliation{Novosibirsk State University,
630090 Novosibirsk, Russia}

\author{S.~I.~Serednyakov}
\affiliation{Budker Institute of Nuclear Physics, 
630090 Novosibirsk, Russia}
\affiliation{Novosibirsk State University,
630090 Novosibirsk, Russia}

\author{Yu.~M.~Shatunov}
\affiliation{Budker Institute of Nuclear Physics, 
630090 Novosibirsk, Russia}
\affiliation{Novosibirsk State University,
630090 Novosibirsk, Russia}

\author{P.~Yu.~Shatunov}
\affiliation{Budker Institute of Nuclear Physics, 
630090 Novosibirsk, Russia}
\affiliation{Novosibirsk State University,
630090 Novosibirsk, Russia}

\author{D.~A.~Shtol}
\affiliation{Budker Institute of Nuclear Physics, 
630090 Novosibirsk, Russia}
\affiliation{Novosibirsk State University,
630090 Novosibirsk, Russia}

\author{Z.~K.~Silagadze}
\affiliation{Budker Institute of Nuclear Physics, 
630090 Novosibirsk, Russia}
\affiliation{Novosibirsk State University,
630090 Novosibirsk, Russia}

\author{A.~N.~Skrinsky}
\affiliation{Budker Institute of Nuclear Physics, 
630090 Novosibirsk, Russia}

\author{I.~K.~Surin}
\affiliation{Budker Institute of Nuclear Physics, 
630090 Novosibirsk, Russia}
\affiliation{Novosibirsk State University,
630090 Novosibirsk, Russia}

\author{Yu.~A.~Tikhonov}
\affiliation{Budker Institute of Nuclear Physics, 
630090 Novosibirsk, Russia}
\affiliation{Novosibirsk State University,
630090 Novosibirsk, Russia}

\author{Yu.~V.~Usov}
\affiliation{Budker Institute of Nuclear Physics, 
630090 Novosibirsk, Russia}
\affiliation{Novosibirsk State University,
630090 Novosibirsk, Russia}

\author{A.~V.~Vasiljev}
\affiliation{Budker Institute of Nuclear Physics,
630090 Novosibirsk, Russia}
\affiliation{Novosibirsk State University,
630090 Novosibirsk, Russia}

\author{I.~M.~Zemlyansky}
\affiliation{Budker Institute of Nuclear Physics, 
630090 Novosibirsk, Russia}
\affiliation{Novosibirsk State University,
630090 Novosibirsk, Russia}


\maketitle

\section{Introduction}
In this paper we continue the study of $e^+e^-$-annihilation into kaon pairs
with the SND detector begun in experiments at the VEPP-2M 
collider~\cite{sndkc,sndkn1,sndkn2}. Data collected
with the upgraded SND detector~\cite{snd1,snd2,snd3,snd4} in experiments
at the VEPP-2000 collider~\cite{vepp} allow to extend the energy range 
under study up to 2 GeV and improve the accuracy of cross section measurements.

One of the goals of the experiments at VEPP-2000 is the measurement of the 
total cross section of $e^+e^-$-annihilation into hadrons necessary
for the Standard Model calculation of the muon anomalous magnetic moment
and the running  electromagnetic coupling. The process
$e^+e^-\to K^+K^-$ studied in this work gives a significant contribution to
the total hadronic cross section in the center-of-mass (c.m.) energy range 
$\sqrt{s}=1-2$ GeV.

The combined analysis of the $e^+e^-\to K^+K^-$ and $e^+e^-\to K_SK_L$ cross
sections and the spectral function in the $\tau^-\to K^-K^0\nu_\tau$ decay 
allows to test the conserved-vector-current hypothesis, as well as to 
separate isovector and isoscalar parts of the $\gamma^\ast\to K\bar{K}$
amplitude. The latter is needed, in particular, to measure 
the branching fractions for the decays of excited
vector states of the $\rho$, $\omega$, $\phi$ families to kaon pairs.

The process $e^+e^-\to K^+K^-$ at energies above the $\phi$-meson resonance
was studied in the OLYA~\cite{olya}, DM1~\cite{dm1}, DM2~\cite{dm2},
SND@VEPP-2M~\cite{sndkc}, and BABAR~\cite{babar} experiments.
The most accurate measurement to date was performed by BABAR using
the initial-state-radiation technique. It should be noted that there are 
significant differences between the SND@VEPP-2M and BABAR measurements 
at $\sqrt{s}<1.4$ GeV, and between the DM2 and BABAR measurements 
at $\sqrt{s}=1.33-1.64$ GeV~\cite{babar}. In this paper,
the $e^+e^-\to K^+K^-$ cross section is measured in the energy range
$\sqrt{s}=1.05 \div 2.00$ GeV with an accuracy not worse than that in the 
BABAR experiment.

The $e^+e^-\to K^+K^-$ study is also useful from the methodical point of view.
The charged kaon identification in the upgraded SND detector is based on 
information from the threshold aerogel Cherenkov counters~\cite{snd3}.
The present analysis is the first work using this
identification system and demonstrates its ability to select events with
charged kaons.
 
\section{Detector and experiment\label{sec_det}}
SND is an universal nonmagnetic detector collecting data at the VEPP-2000 
$e^+e^-$ collider. 
The main part of the detector is a three-layer electromagnetic
calorimeter~\cite{snd1} consisting of 1640 NaI(Tl) crystals. The total 
thickness of the calorimeter is 13.4 radiation lengths. Its energy
resolution for photons is $\sigma_E/E = 4.2\% /\sqrt[4\,]{E(\mathrm{GeV})}$,
and the angular resolution is $\sigma_{\phi},\sigma_{\theta}\simeq1.5^\circ$.
The solid-angle coverage of the calorimeter is about 95\% of the $4\pi$. 
Inside the calorimeter, a nine-layer drift chamber~\cite{snd2} is installed,
which is used for measurement of directions and production points of charged particles.
Charged particle identification is based on information from the system of 
threshold aerogel Cherenkov counters (ACC)~\cite{snd3}.
It consists of 9 counters, which form a cylinder located
directly behind the drift chamber. The thickness of the aerogel is about 30 mm.
The counters cover the polar angle range $50^\circ<\theta<132^\circ$.
The Cherenkov light is collected using wavelength shifters
located inside the aerogel radiator. 
In the data analysis, the coordinates of the particle entrance to ACC are 
calculated. Information from ACC is used only if the particle track 
extrapolates to the ACC active area that excludes the regions of shifters 
and gaps between counters. The active area is 81\% of the ACC area.
There are two ACC options, with a refractive index of 1.05 and 1.13.
At energies above the threshold of kaons production ACC with 
the higher refractive index is used, and a kaon is identified by the 
requirement of no Cherenkov signal in ACC. For pions, the threshold momentum 
is about 265 MeV/$c$.

In this paper, data with an integrated luminosity
of 34.6~pb$^{-1}$ are analyzed, which were recorded in several scans of the 
c.m. energy interval from 1.05 to 2.00 GeV in 2011 and 2012.

During the experiments, the beam energy was monitored using measurements of the
magnetic field in the collider bending magnets. For absolute calibration of 
the collider energy, a scan of the $\phi(1020)$ resonance and its mass 
measurement were performed. In 2012, the energy was measured in several energy
points near 2 GeV using the back-scattering-laser-light system~\cite{COMPTON}.
The absolute energy measurements were used to calibrate the momentum 
measurement in the CMD-3 detector, which collected data at VEPP-2000 in 
parallel with SND. The c.m. energies for all scan points were then determined 
with an accuracy of 2--6 MeV using average momentum of Bhabha and
$e^+e^-\to p\bar{p}$ events~\cite{BEAM1,BEAM2}.

\section{\boldmath Selection of $e^+e^-\to K^+ K^-$ events}
The $e^+e^-\to K^+ K^-$ events are detected as a pair of back-to-back
(collinear) charged particles. Events may contain extra charged tracks 
originating from $\delta$-electrons or beam background, and spurious photons
originating from beam background and kaon nuclear interaction in 
the calorimeter. We select events with at least two reconstructed charged 
particles. Two of them with highest energy deposition in the calorimeter
must satisfy the following requirements:
\begin{itemize}
\item[--] the distance between the track and the beam axis $|d_{i}| < 0.25$ cm;
\item[--] the difference between $z$ coordinates of the interaction point
and the point at the track closest to the beam axis $|z_{i}| < 7$ cm;
\item[--] $|z_1-z_2| < 1$ cm;
\item[--] $|\Delta\phi|< 10^\circ$ for $\sqrt{s}<1.09$ GeV,\\ 
$|\Delta\phi|< 5^\circ$ for $1.09 < \sqrt{s} < 1.20$ GeV,\\
$|\Delta\phi|< 3^\circ$ for $\sqrt{s}> 1.2$ GeV,\\
where $\Delta\phi=|\phi_1-\phi_2|-180^\circ$, 
and $\phi_i$ is the track azimuthal angle;
\item[--] $|\Delta\theta|< 10^\circ$ for $\sqrt{s} < 1.2$ GeV,\\
$|\Delta\theta|< 7^\circ$ for $\sqrt{s}> 1.2$ GeV,\\
where 
$\Delta\theta=\theta_1+\theta_2-180^\circ$, and $\theta_i$ is 
the track polar angle;
\item[--] one of the particles extrapolates to the ACC active area
and does not produce any signal in ACC.
\end{itemize}
At $\sqrt{s} < 1.2$ GeV, a significant suppression of background processes
with electrons, muons and pions in the final state can be achieved 
using ionization energy loss $(dE/dx)$ measurements in the drift chamber. 
The following condition on the sum of $dE/dx$ of two particles is applied:
$(dE/dx)_1+(dE/dx)_2 > k (dE/dx)_e$, where $(dE/dx)_e$ is the average $dE/dx$
for electrons, and the coefficient $k$ is equal to 3 for $\sqrt{s} < 1.1$, 
and 2.5 for $1.1 < \sqrt{s} < 1.2$ GeV.

At $\sqrt{s} > 1.9$ GeV it is required that $dE/dx$ of one of the 
charged particles do not exceed $1.5(dE/dx)_e$. This condition is used to 
suppress the background from $e^+e^-\to p\bar{p}$ events.

\section{Background subtraction\label{bkg}}
Background events can be divided into two groups, collinear and noncollinear.
The first group includes two-body processes $e^+e^-\to e^+e^-$,
$\mu^+\mu^-$, $\pi^+\pi^-$ and $p\bar{p}$, as well as events with cosmic
muons passing near the interaction point. The second group contains mainly 
multibody processes with two charged particles: $e^+e^-\to \pi^+\pi^-\pi^0$,
$\pi^+\pi^-\pi^0\pi^0$, $K^+K^-\pi^0$, etc.
The process $e^+e^-\to \phi\gamma \to K^+ K^- \gamma$, where the photon is 
emitted from the initial state, also contributes to the second group. 
This process is strongly suppressed by the condition on $\Delta\theta$.
Its contribution is significant only near $\phi$-meson resonance, at 
$\sqrt{s} < 1.06$ GeV, and in the narrow region $1.75 < \sqrt{s} < 1.80$ GeV,
where the $e^+e^-\to K^+ K^- \gamma$ cross section is very small.

The background subtraction is performed in two stages.
At the first stage the number of background noncollinear events is determined.
The collinear background is subtracted at the second stage.

\subsection{Noncollinear background}
\begin{figure}
\includegraphics[width=.45\textwidth]{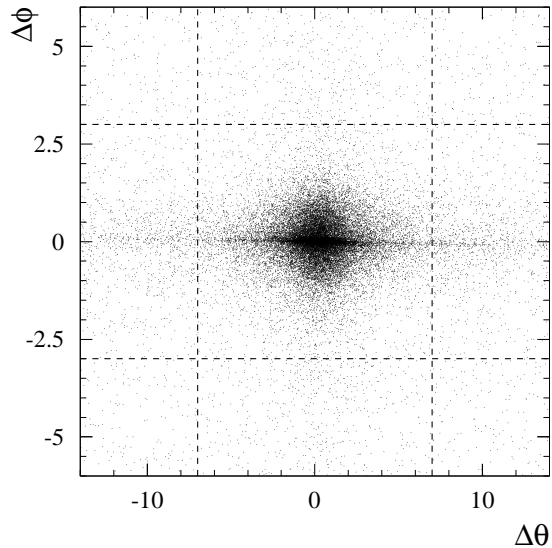}
\caption{\label{dfdt}The two-dimensional $\Delta \phi$ versus $\Delta \theta$ 
distribution for selected data events with $\sqrt{s}=1.4-1.6$ GeV. The dashed 
lines indicate the boundaries of the conditions on 
$\Delta \phi$ and $\Delta \theta$. The central box is 
a signal region corresponding the standard selection criteria for
$e^+e^-\to K^+ K^-$ candidates.}
\end{figure}
\begin{figure}
\includegraphics[width=.45\textwidth]{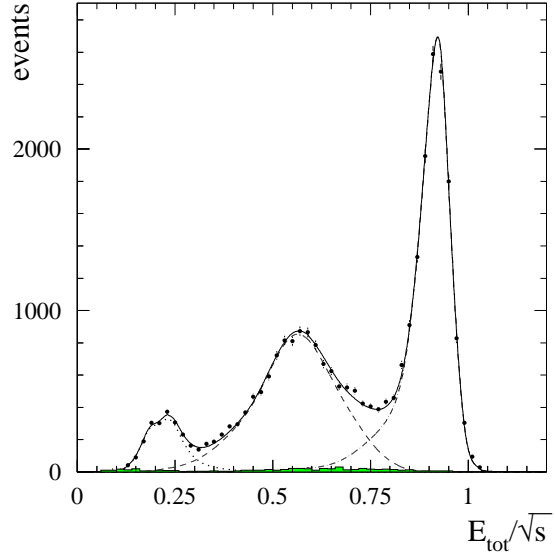}
\caption{\label{spect1}
The distribution of the normalized total energy deposition in the calorimeter
for selected data events with $\sqrt{s}=1.4-1.6$ GeV (points with error bars).
The shaded histogram represents the noncollinear background. The 
dotted, dashed, and dash-dotted curves 
show the contribution of $e^+e^-\to \mu^+\mu^-$ and cosmic events, 
$e^+e^-\to K^+ K^-$ events, and $e^+e^-\to e^+e^-$ events, respectively.
The solid curve is the sum of these three components.}
\end{figure}
The noncollinear background is estimated from the
two-dimensional distribution of $\Delta \phi$ and $\Delta \theta$.
Such a distribution for selected data events from the 
energy range $\sqrt{s}=1.4-1.6$ GeV is shown in Fig.~\ref{dfdt}. 
The dashed lines indicate the boundaries of the conditions on
$\Delta \phi$ and $\Delta \theta$. The central box in the plot is a signal 
region corresponding to the standard selection criteria for $e^+e^-\to K^+ K^-$ 
candidates. The peak of signal and background collinear processes is clearly 
seen in the center of the signal region. Noncollinear processes is expected 
to have a flat $\Delta \phi$ vs $\Delta \theta$ distribution. The number of 
noncollinear background events in the signal region ($n_{\rm bkg}$)
is estimated from the number of events in four regions located in the corners
of the two-dimensional plot ($n^\prime_{\rm bkg}$) as
$n_{\rm bkg}=\alpha_{\rm bkg}n^\prime_{\rm bkg}$. The value of the 
$\alpha_{\rm bkg}$ coefficient is estimated from Monte Carlo (MC) simulation 
of the main noncollinear background processes $e^+e^-\to 3\pi$, $e^+e^-\to 4\pi$, 
$e^+e^-\to K^+ K^-\pi^0$, $e^+e^-\to K^+ K^-\eta$ and found to
be equal to unity with 10\% uncertanty.
Figure~\ref{spect1} shows the distribution of the normalized total energy 
deposition in the calorimeter $E_{tot}/\sqrt{s}$ for data events from the
signal region. The shaded histogram shows the contribution of noncollinear
background events estimated using the procedure described above.
In further analysis, the noncollinear background is subtracted 
from the number of events in the signal region. Its fraction changes slowly 
from 3\% at 1.1 GeV to 5\% at 1.65 GeV, and then increases up to 40\% at 
$\sqrt{s} > 1.8$ GeV.

\subsection{Collinear background}
\begin{figure}
\includegraphics[width=.45\textwidth]{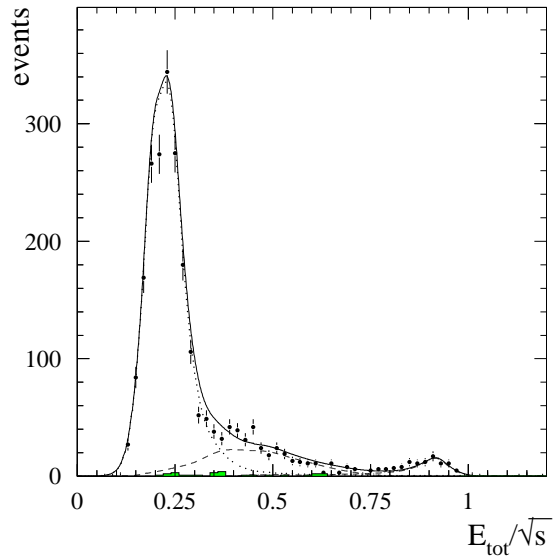}
\caption{\label{spect2}
The distribution of the normalized total energy deposition in the calorimeter
for data events with $\sqrt{s}=1.4-1.6$ GeV selected with the additional requirement that 
the muon system fires (points with error bars). The shaded histogram 
represents the noncollinear background. The descriptions of the curves
are given in the caption of Fig.~\ref{spect1}.}
\end{figure}
\begin{figure}
\includegraphics[width=.45\textwidth]{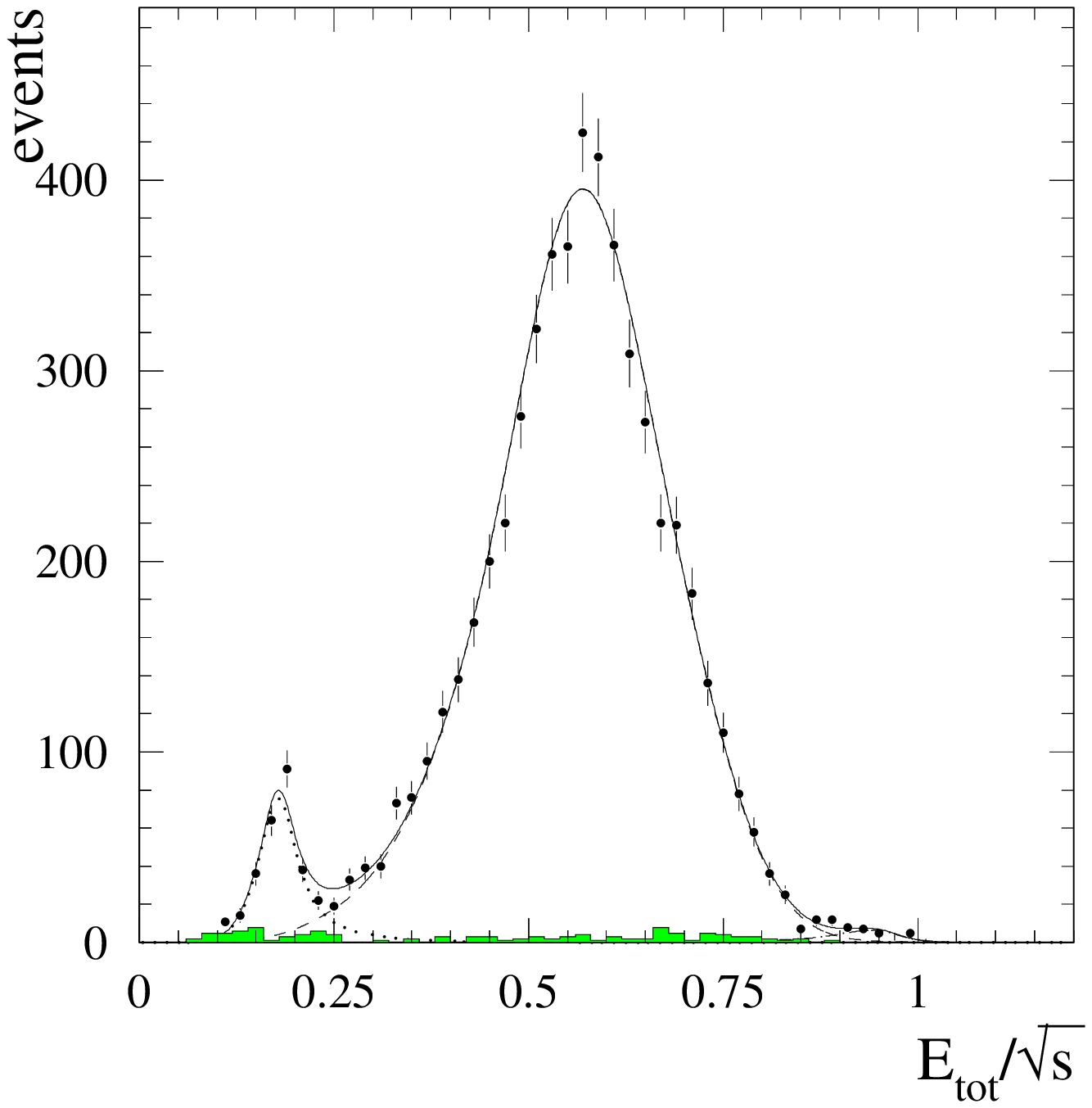}
\caption{\label{spect3}
The distribution of the normalized total energy deposition in the calorimeter
for data events with $\sqrt{s}=1.4-1.6$ GeV in which the condition of no ACC signal is applied to
both particles (points with error bars). The shaded histogram 
represents the noncollinear background. The descriptions of the curves
are given in the caption of Fig.~\ref{spect1}.}
\end{figure}
The contributions of the process under study (middle peak) and background 
processes, $e^+e^-\to \mu^+\mu^-$ plus cosmic muons (left peak) and $e^+e^-\to e^+e^-$
(right peak), are clearly seen in the $E_{tot}/\sqrt{s}$ distribution shown 
in Fig.~\ref{spect1}. It should be noted that the requirement of 
no ACC signal for one of the charged particles suppresses 
$e^+e^-\to e^+e^-$ and $e^+e^-\to \mu^+\mu^-$ events by a factor of about 300.
Background from the $e^+e^-\to p\bar{p}$ process is completely rejected by 
the selection criteria used, in particular, the requirement on $dE/dx$.

The $E_{tot}/\sqrt{s}$ distribution for $e^+e^-\to \pi^+\pi^-$ events is
close to that for kaons. Their fraction under the kaon peak calculated using 
MC simulation decreases from 5\% at 1.2 GeV to 0.1\% at $\sqrt{s}=1.6$ GeV,
and then increases reaching about 8\% at 1.8 GeV.
The calculation uses the $e^+e^-\to \pi^+\pi^-$ cross section measured in
Ref.~\cite{pipi}.

The accuracy of the $e^+e^-\to \pi^+\pi^-$ simulation and, in particular, the 
probability for a pion to not produce a signal in ACC are tested in the energy 
range $1.05 \geq \sqrt{s} \leq 1.15$ GeV, where the cross section of this 
process is relatively high. The standard criteria are applied to select 
$e^+e^-\to \pi^+\pi^-$ events, except the condition on $(dE/dx)$, which is
inverted to suppress events with kaons: $(dE/dx)_1+(dE/dx)_2 < 2 (dE/dx)_e$.
Events of the $e^+e^-\to e^+e^-$ process are suppressed by the requirement 
$E_{tot}/\sqrt{s} < 0.7$, while $e^+e^-\to \mu^+\mu^-$ and cosmic background is
rejected by the muon-system veto and the condition that the energy deposition
in the upper half of the calorimeter is outside the range of the peak from 
minimum-ionizing particles. The number of selected $e^+e^-\to \pi^+\pi^-$
events coincides with the number expected from simulation within 10\%. This 
value is taken as an estimate of the systematic uncertainty 
on the pion background calculation.

To determine the number of kaon events, the $E_{tot}/\sqrt{s}$ distribution
is fitted by a sum of muon, kaon and electron distributions. 
The distributions for background processes are obtained from data. 
A practically pure sample of $e^+e^-\to e^+e^-$ events is selected by the
requirement that one of the particles in an event passes through the ACC active 
area and produces a Cherenkov signal. The muon distribution is obtained using
events selected with the requirement that the muon system fires. The 
$E_{tot}/\sqrt{s}$ distribution for such events is shown in Fig.~\ref{spect2}.
It is seen that the muon peak survives, while kaon and electron events are
strongly suppressed. To study the distribution for cosmic muons, the condition
on the parameters $d_i$ is modified: $0.25 < |d_{1,2}| < 0.5$. This
allows to reject $e^+e^-$ annihilation events and obtain a pure spectrum 
of cosmic muons. It is found that the muon distribution consist of two
components. One of the components, with higher energy deposition, contains
$e^+e^-\to \mu^+\mu^-$ events and events with energetic cosmic muons. 
The other component contains cosmic muons with momentum less than
the threshold momentum for muons in ACC. This second component is seen in 
Fig.~\ref{spect3}, which shows the $E_{tot}/\sqrt{s}$ distribution for events,
in which the condition of no ACC signal is applied to both particles.

The kaon distribution is obtained using simulation.
To take into account a possible inaccuracy in simulation of detector response,
the distribution is convolved with a Gaussian distribution, the parameters
which (a shift of the peak position $\delta x$ and sigma $\sigma_x$, where
$x\equiv E_{tot}/\sqrt{s}$) are determined from the fit to data. 
The fit results for events from the energy range $\sqrt{s}=1.4-1.6$ selected
with different conditions are shown in Figs.~\ref{spect1}, \ref{spect2} and
\ref{spect3}. It is seen that the fitted curves describe data well. For this 
range the parameters $\delta x$ and $\sigma_x$ are found 
to be $-0.019\pm0.002$ and $0.034\pm0.006$, respectively. 

The fitted numbers of kaon events with subtracted $e^+e^-\to \pi^+\pi^-$
contribution are listed in Tables~\ref{table1} and \ref{table2} for
the 2011 and 2012 scans, respectively.
\begin{table}
\caption{\label{table1}Data for the 2011 scan.
The c.m. energy ($\sqrt{s}$), integrated luminosity ($L$),
number of selected $e^+e^-\to K^+K^-$ events ($N_{exp}$), detection
efficiency ($\varepsilon_0$), 
radiative correction factor ($1+\delta$), and $e^+e^-\to K^+K^-$ Born cross 
section ($\sigma_0$). For the number of events,
only the statistical uncertainty  is quoted. For the cross section, the
first error is statistical, the second is systematic.}
\begin{ruledtabular}
\begin{tabular}{cccccc}
$\sqrt{s}$, GeV&$L$, nb$^{-1}$&$N_{exp}$&$\varepsilon_0$&$1+\delta$&$\sigma_0$, nb \\
\hline
   1.047 &    426 & $  3975\pm  63$ &  0.229 &  1.126 & $36.243\pm 0.592\pm 0.471$ \\
   1.075 &    566 & $  3744\pm  61$ &  0.400 &  0.895 & $18.483\pm 0.305\pm 0.240$ \\
   1.097 &    568 & $  3436\pm  59$ &  0.487 &  0.876 & $14.184\pm 0.244\pm 0.184$ \\
   1.124 &    550 & $  3292\pm  58$ &  0.594 &  0.890 & $11.307\pm 0.199\pm 0.147$ \\
   1.151 &    499 & $  2460\pm  50$ &  0.584 &  0.896 & $ 9.415\pm 0.192\pm 0.122$ \\
   1.174 &    557 & $  2917\pm  54$ &  0.638 &  0.898 & $ 9.148\pm 0.171\pm 0.119$ \\
   1.196 &    566 & $  2441\pm  48$ &  0.618 &  0.903 & $ 7.737\pm 0.153\pm 0.101$ \\
   1.223 &    575 & $  2350\pm  56$ &  0.639 &  0.880 & $ 7.269\pm 0.175\pm 0.104$ \\
   1.245 &    480 & $  1747\pm  49$ &  0.624 &  0.879 & $ 6.624\pm 0.187\pm 0.092$ \\
   1.273 &    513 & $  1928\pm  50$ &  0.654 &  0.881 & $ 6.518\pm 0.170\pm 0.089$ \\
   1.295 &    497 & $  1680\pm  52$ &  0.643 &  0.881 & $ 5.964\pm 0.184\pm 0.081$ \\
   1.323 &    565 & $  1900\pm  51$ &  0.666 &  0.883 & $ 5.718\pm 0.154\pm 0.077$ \\
   1.344 &    598 & $  1801\pm  52$ &  0.660 &  0.885 & $ 5.160\pm 0.148\pm 0.069$ \\
   1.374 &    626 & $  1971\pm  49$ &  0.668 &  0.888 & $ 5.308\pm 0.132\pm 0.072$ \\
   1.394 &    624 & $  1834\pm  48$ &  0.661 &  0.889 & $ 5.001\pm 0.131\pm 0.067$ \\
   1.423 &    588 & $  1639\pm  43$ &  0.683 &  0.891 & $ 4.579\pm 0.121\pm 0.063$ \\
   1.443 &    473 & $  1200\pm  39$ &  0.668 &  0.893 & $ 4.254\pm 0.137\pm 0.057$ \\
   1.471 &    620 & $  1551\pm  42$ &  0.686 &  0.891 & $ 4.093\pm 0.111\pm 0.056$ \\
   1.494 &    754 & $  1648\pm  50$ &  0.672 &  0.892 & $ 3.646\pm 0.110\pm 0.049$ \\
   1.522 &    508 & $  1138\pm  38$ &  0.684 &  0.889 & $ 3.679\pm 0.124\pm 0.050$ \\
   1.543 &    578 & $  1159\pm  42$ &  0.668 &  0.889 & $ 3.382\pm 0.122\pm 0.045$ \\
   1.572 &    533 & $  1140\pm  39$ &  0.684 &  0.889 & $ 3.518\pm 0.121\pm 0.048$ \\
   1.594 &    462 & $   959\pm  41$ &  0.667 &  0.888 & $ 3.507\pm 0.152\pm 0.047$ \\
   1.623 &    545 & $  1010\pm  34$ &  0.684 &  0.898 & $ 3.022\pm 0.102\pm 0.043$ \\
   1.643 &    499 & $   846\pm  32$ &  0.662 &  0.911 & $ 2.815\pm 0.106\pm 0.039$ \\
   1.669 &    483 & $   663\pm  28$ &  0.679 &  0.937 & $ 2.155\pm 0.091\pm 0.031$ \\
   1.693 &    490 & $   494\pm  25$ &  0.668 &  0.956 & $ 1.570\pm 0.081\pm 0.023$ \\
   1.723 &    539 & $   349\pm  25$ &  0.682 &  0.976 & $ 0.968\pm 0.069\pm 0.020$ \\
   1.742 &    529 & $   224\pm  18$ &  0.662 &  1.014 & $ 0.633\pm 0.051\pm 0.013$ \\
   1.774 &    485 & $   111\pm  13$ &  0.683 &  1.100 & $ 0.310\pm 0.036\pm 0.009$ \\
   1.793 &    412 & $    50\pm  10$ &  0.667 &  1.084 & $ 0.170\pm 0.035\pm 0.008$ \\
   1.826 &    529 & $    74\pm  12$ &  0.685 &  0.957 & $ 0.215\pm 0.034\pm 0.012$ \\
   1.849 &    438 & $    44\pm  10$ &  0.654 &  0.895 & $ 0.171\pm 0.038\pm 0.020$ \\
   1.871 &    669 & $   116\pm  15$ &  0.683 &  0.871 & $ 0.291\pm 0.036\pm 0.017$ \\
   1.893 &    624 & $   125\pm  15$ &  0.668 &  0.867 & $ 0.345\pm 0.040\pm 0.016$ \\
   1.901 &    494 & $    96\pm  14$ &  0.650 &  0.867 & $ 0.343\pm 0.049\pm 0.013$ \\
   1.927 &    626 & $   111\pm  15$ &  0.644 &  0.872 & $ 0.316\pm 0.042\pm 0.014$ \\
   1.953 &    330 & $    66\pm  11$ &  0.637 &  0.878 & $ 0.357\pm 0.061\pm 0.015$ \\
   1.978 &    449 & $    85\pm  14$ &  0.642 &  0.886 & $ 0.332\pm 0.055\pm 0.017$ \\
   2.005 &    582 & $   122\pm  16$ &  0.641 &  0.893 & $ 0.367\pm 0.048\pm 0.016$ \\
\end{tabular}
\end{ruledtabular}
\end{table}
\clearpage
\begin{table*}
\caption{\label{table2}Data for the 2012 scan.
The c.m. energy ($\sqrt{s}$), integrated luminosity ($L$),
number of selected $e^+e^-\to K^+K^-$ events ($N_{exp}$), detection
efficiency ($\varepsilon_0$), radiative correction factor ($1+\delta$),
and $e^+e^-\to K^+K^-$ Born cross section ($\sigma_0$). For number of events,
only the statistical uncertainty  is quoted. For the cross section, the
first error is statistical, the second is systematic.}
\begin{ruledtabular}
\begin{tabular}{cccccc}
$\sqrt{s}$, GeV& $L$, nb$^{-1}$&$N_{exp}$&$\varepsilon_0$&$1+\delta$&$\sigma_0$, nb \\
\hline
   1.277 &    763 & $  2795\pm  62$ &  0.653 &  0.882 & $ 6.358\pm 0.142\pm 0.105$ \\
   1.357 &    845 & $  2676\pm  61$ &  0.670 &  0.886 & $ 5.339\pm 0.122\pm 0.089$ \\
   1.435 &   1032 & $  2556\pm  61$ &  0.675 &  0.892 & $ 4.114\pm 0.099\pm 0.069$ \\
   1.515 &    940 & $  2031\pm  54$ &  0.678 &  0.891 & $ 3.574\pm 0.094\pm 0.060$ \\
   1.595 &    822 & $  1604\pm  44$ &  0.666 &  0.888 & $ 3.302\pm 0.090\pm 0.055$ \\
   1.674 &    914 & $  1044\pm  36$ &  0.682 &  0.944 & $ 1.770\pm 0.062\pm 0.031$ \\
   1.716 &    512 & $   373\pm  25$ &  0.673 &  0.967 & $ 1.115\pm 0.075\pm 0.025$ \\
   1.758 &    804 & $   282\pm  21$ &  0.658 &  1.068 & $ 0.503\pm 0.038\pm 0.014$ \\
   1.798 &   1012 & $   136\pm  18$ &  0.683 &  1.069 & $ 0.187\pm 0.025\pm 0.016$ \\
   1.840 &    568 & $   103\pm  15$ &  0.676 &  0.915 & $ 0.293\pm 0.041\pm 0.014$ \\
   1.874 &    936 & $   152\pm  18$ &  0.675 &  0.870 & $ 0.276\pm 0.033\pm 0.016$ \\
   1.903 &    962 & $   162\pm  18$ &  0.677 &  0.867 & $ 0.287\pm 0.033\pm 0.017$ \\
   1.926 &    680 & $   148\pm  18$ &  0.679 &  0.872 & $ 0.367\pm 0.044\pm 0.016$ \\
   1.945 &    929 & $   145\pm  19$ &  0.676 &  0.876 & $ 0.263\pm 0.034\pm 0.015$ \\
   1.967 &    755 & $   147\pm  18$ &  0.666 &  0.885 & $ 0.331\pm 0.041\pm 0.015$ \\
   1.989 &    641 & $   131\pm  17$ &  0.666 &  0.889 & $ 0.346\pm 0.046\pm 0.015$ \\
\end{tabular}
\end{ruledtabular}
\end{table*}

\section{Detection efficiency\label{detef}}
\begin{figure}
\includegraphics[width=.45\textwidth]{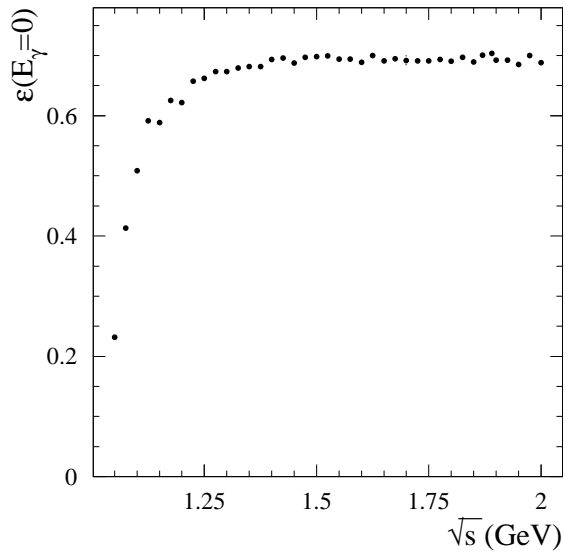}
\caption{\label{eff0}
The energy dependence of the detection efficiency for $e^+e^-\to K^+ K^-$ 
events with $E_\gamma=0$.}
\end{figure}
\begin{figure}
\includegraphics[width=.45\textwidth]{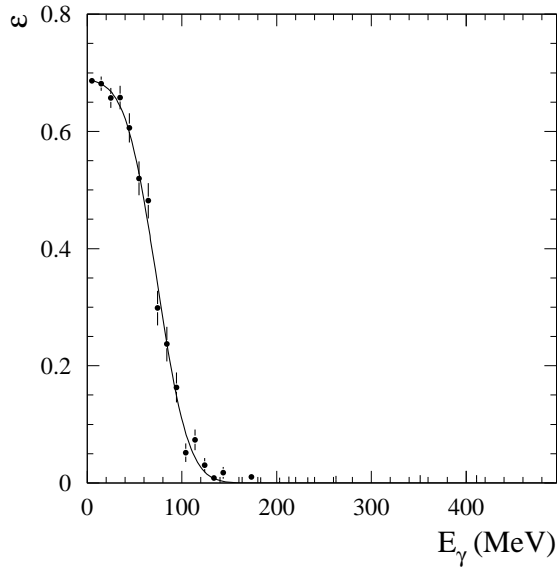}
\caption{\label{eff2}
The dependence of the detection efficiency on the energy of the photon
emitted from the initial state at $\sqrt{s}=1.6$ GeV.}
\end{figure}
The detection efficiency for the process under study is determined using MC
simulation. The simulation takes into account extra photon radiation from 
the initial state~\cite{RadCor, RadCor2}. The detection efficiency 
($\varepsilon$) is calculated as a function of $\sqrt{s}$ and the energy of 
the photon ($E_\gamma$) emitted by the initial particles. The energy dependence 
of the detection efficiency for $e^+e^-\to K^+ K^-$ events with $E_\gamma=0$
is shown in Fig.~\ref{eff0}. The decrease of the efficiency when
the energy approaches the reaction threshold is explained by an increase 
of the fraction of events with a kaon decayed before ACC. A nonmonotonic
behavior of the detection efficiency as a function of energy is due to
variation of experimental conditions during data taking, in particular,
the number of dead detector channels and the level of beam background.
A typical dependence of the detection efficiency on $E_\gamma$ is shown in
Fig.~\ref{eff2}.

The efficiency determined from MC simulation is corrected to take into account
a difference between data and simulation in detector response and the
absence of the final state radiation (FSR) in the $e^+e^-\to K^+ K^-$ simulation.
Corrections associated with inaccuracy of detector simulation is described
by three correction factors: for kinematic selection criteria ($c_{\rm kin}$),
e.g., the conditions on $\Delta\theta$ and $\Delta\phi$, for the 
geometric condition that one of the particles must pass through the active ACC
area ($c_{\rm geo}$), and for the kaon identification condition ($c_{\rm id}$).
To measure the corrections
of the first group, the condition on the parameter under study is loosened. 
Tighter conditions may be applied on other kinematic parameters.
The correction factor is calculated as follows:
\begin{equation}
c_i=\frac{n/n^\prime}{m/m^\prime},
\label{noncollbkg}
\end{equation}
where $n$ and $m$ are the numbers of events in data and simulation obtained
with the standard condition on the parameter under study, and $n^\prime$ and 
$m^\prime$ are the same numbers with a looser condition.
The total correction factor for the kinematic conditions is calculated
as a product of the factors obtained for each of the condition and is found 
to be $c_{\rm kin} =1.008\pm0.002$.

The identification correction is determined as follows.
The cross section for $e^+e^-\to K^+ K^-$ is measured for two sets of 
selection criteria. In the first set it is required that both particles 
pass through the active ACC area, but only one of them is identified as a kaon
(no signal in ACC). In the second set it is required that both particles
pass through the active ACC area and are identified as kaons.
Since the geometric conditions are identical in the both selections,
the ratio of the cross sections is equal to the value of the identification 
correction factor. It is found that $c_{\rm id}$ is independent of energy and
is equal to $1.003\pm0.007$ for the 2011 scan and $1.004\pm0.012$ for the 2012
scan. The closeness of the obtained correction factors to unity indicates that
our simulation reproduces the ACC response for kaons well. 

The geometric correction is estimated using $e^+e^-\to e^+e^-$ events.
The fractions of events, in which one of the the 
particle tracks extrapolates to the ACC active area, are determined 
in data and simulation. From their ratio the geometric correction factor is 
obtained to be $c_{\rm geo} = 1.0017\pm0.0004$ for the 2011 scan and 
$c_{\rm geo} = 0.9974\pm0.0007$ for the 2012 scan.

As was mentioned above, our signal simulation does not include FSR and 
therefore reproduces the $\Delta\theta$ and $\Delta\phi$ distributions 
incorrectly. 
In particular, some fraction of FSR events falls
into the corner regions of the two-dimensional $\Delta\phi$ vs $\Delta\theta$
distribution in Fig.~\ref{dfdt}, which are used to estimate the
noncollinear background. The effect of FSR on the detection efficiency is
studied with the event generator based on Ref.~\cite{arbuzov}, which includes 
both initial and final state radiation. Using MC simulation at the generator
level the detection efficiencies calculated with and without FSR are compared. 
The inclusion of FSR reduces the efficiency by 0.1\% at 1.1 GeV, 0.7\% at 1.5
GeV and 1.3\% at 2.0 GeV.

The corrected values of the detection efficiency 
$\varepsilon_0\equiv\varepsilon(E_\gamma=0)$
are listed in Tables~\ref{table1} and \ref{table2}
for 2011 and 2012, respectively.
The uncertainty of the total correction factor is 0.7\% for 2011 and 
1.2\% for 2012. These values are taken as an estimate of the systematic 
uncertainties on the detection efficiency.

\section{Born cross section}
The experimentally observed cross section of the process under study 
$\sigma_{\rm vis}$ is related to the Born cross section $\sigma_{0}$ by the 
formula:
\begin{equation}
\sigma_{\rm vis}(\sqrt{s})=\int\limits^{z_{\rm max}}_{0} dz\sigma_{0}
(\sqrt{s(1-z)})F(z,s)\varepsilon(\sqrt{s},z),
\label{bornsec}
\end{equation}
where $F(z,s)$ is a function describing the probability to emit extra photons
with the total energy $z\sqrt{s}/2$ from the initial state~\cite{RadCor},
$\varepsilon(\sqrt{s},z)$ is the detection efficiency, and 
$z_{\rm max}=1-4m_K^2/s$. The following procedure is used to determine the 
experimental values of the Born cross section. The measured cross section
$\sigma_{{\rm vis},i}={N_{{\rm exp},i}}/{L_i}$,
where $N_{{\rm exp},i}$ is the number of selected events with subtracted 
background and $L_i$ is the integrated luminosity for $i$-th energy point, 
is fitted by Eq.~(\ref{bornsec}) with a theoretical model for the Born cross 
section. As a result of the fit, parameters of the model are determined, and
the radiation correction factor is calculated as
$1+\delta(s)=\sigma_{\rm vis}(s)/(\varepsilon_0(s)\sigma_{0}(s))$, where
$\varepsilon_0(s)\equiv\varepsilon(s,z=0)$.
The experimental values of the Born cross sections are then obtained as
\begin{equation}
\sigma_{0,i}=\frac{\sigma_{{\rm vis},i}}{\varepsilon_0(s_i)(1+\delta(s_i))}.
\label{secborni}
\end{equation}
\begin{figure}
\includegraphics*[width=0.9\textwidth, trim=0mm 20mm 0mm 0mm]{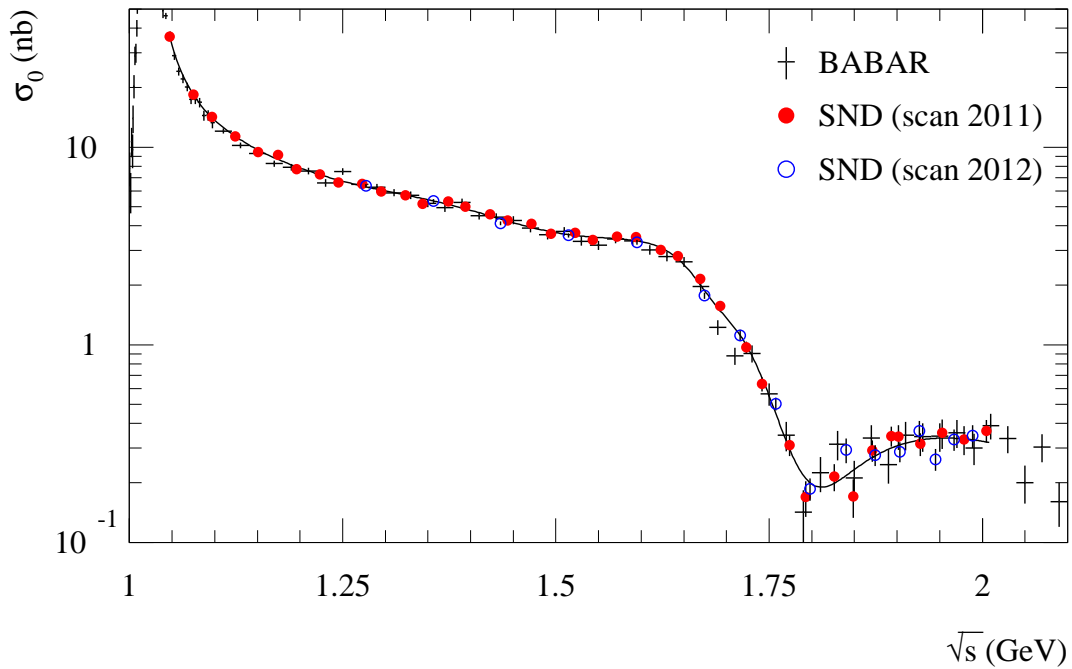}
\caption{\label{crsecb}
The Born cross section for the process $e^+e^-\to K^+K^-$  measured in this
work and in the BABAR experiment~\cite{babar}. 
The curve is the result of the fit described in the text.}
\end{figure}

The vector meson dominance model is used to describe the Born 
$e^+e^-\to K^+K^-$ cross section~\cite{vmd}: 
\begin{eqnarray}
\sigma_{0}(s)&=& \frac{\pi\alpha^2\beta^3}{3s}|F_K(s)|^2(1+C_{\rm FS}),\\
F_K(s)&=&\sum_{V=\rho,\omega,\phi,{\ldots}}a_Ve^{i\phi_V}\frac{m_V^2}
{m_V^2-s-im_V \Gamma_V(s)},
\end{eqnarray}
where $\alpha$ is the fine structure constant, $\beta=\sqrt{1-4m_K^2/s}$,
$C_{\rm FS}$ is a factor describing radiation corrections to the final 
state~\cite{cfs}, which varies from about 3\% at $\sqrt{s}=1.05$ GeV to 0.9\% 
at $\sqrt{s}=2.0$ GeV. The charged kaon electromagnetic form factor $F_K(s)$ is
a sum of the contributions of the $\rho$, $\omega$ and $\phi$ resonances and 
their excited states. The masses ($m_V$), widths ($\Gamma_V$), amplitudes 
($a_V$), and the relative phases ($\phi_V$) of the $\rho$, $\omega$ and $\phi$
are fixed using Particle Data Group (PDG) data~\cite{PDG} and SU(3) relations.
For the masses and widths of excited states, PDG values are taken, while their
amplitudes and phases are free fit parameters. The model describes data 
reasonably well 
($\chi^2/ndf=65/46$). The fit result is shown in Fig.~\ref{crsecb}. 
The values of the Born cross section calculated using Eq.~(\ref{secborni})
together with the values of the radiation correction factors
are listed in Tables~\ref{table1} and ~\ref{table2}.

\section{Systematic errors}
\begin{table}
\caption{\label{table3}
The average over the energy intervals systematic uncertainties (\%) on the
measured $e^+e^-\to K^+K^-$ cross section.}
\begin{ruledtabular}
\begin{tabular}{lcccc}
 source                    & \multicolumn{2}{c}{2011} & \multicolumn{2}{c}{2012} \\
                           & $\sqrt{s}<1.8$ GeV & $\sqrt{s}>1.8$ GeV & $\sqrt{s}<1.8$ GeV & $\sqrt{s}>1.8$ GeV \\
\hline
 Luminosity                & 1.0   & 1.0   & 1.0   & 1.0   \\
 Detection efficiency      & 0.7 & 0.7 & 1.2 & 1.2 \\
 Background subtraction    & 0.7 & 4.1 & 0.7 & 4.1 \\
 Nuclear interaction       & 0.1 & 0.1 & 0.1 & 0.1 \\
 Radiative correction      & 0.1 & 0.1 & 0.1 & 0.1 \\
\hline
 Total                     & 1.4 & 4.3 & 1.7 & 4.4 \\
\end{tabular}
\end{ruledtabular}
\end{table}
The sources of systematic uncertainties on the measured cross section are 
listed in Table~\ref{table3}.

The integrated luminosity is measured on events of the processes 
$e^+e^-\to e^+e^-$ and $e^+e^-\to\gamma\gamma$, the cross sections for which
are known with an accuracy better than 1\%. The relative difference between 
these two measurements is independent of the c.m. energy and does not exceed
1\%. This value is taken as an estimate of the systematic uncertainty on the 
luminosity measurement.

The uncertainty on the detection efficiency is discussed in Sec.~\ref{detef}.
The systematic uncertainties associated with the subtraction of the 
noncollinear and $e^+e^-\to \pi^+\pi^-$ backgrounds are discussed in 
Sec.~\ref{bkg}. Accuracy of the subtraction of the muon and electron backgrounds
in the fit to the $E_{tot}/\sqrt{s}$ distribution is tested by the comparison
of the cross sections measured on events with one and two identified kaons.
The ratio of the cross sections is described by the coefficient $c_{\rm id}$.
Since both electron and muon contributions to the $E_{tot}/\sqrt{s}$ 
spectra with these two selections are strongly different (see Fig.~\ref{spect1}
and \ref{spect3}) and the $c_{\rm id}$ value is consistent with unity,
we do not introduce an additional systematic uncertainty associated with 
the fit to the $E_{tot}/\sqrt{s}$ distribution.

A part of kaon events is lost due to the kaon nuclear interaction
in the material before the drift chamber. Its thickness is about $0.5\%$ of
the nuclear interaction length. Taking into account that the cross sections
for the charged kaon nuclear interactions are well known, we estimate that 
the systematic uncertainty associated with the kaon nuclear interaction
does not exceed 0.1\%. The theoretical uncertainty on the radiative correction
calculation is less than 0.1\%~\cite{RadCor}.

We assume that all the sources of systematic uncertainties are independent 
and add them in quadrature. The resulting total systematic uncertainty is
listed in Table~\ref{table3}.

\section{Summary}
In this paper, the $e^+e^-\to K^+K^-$ cross section has been measured
in the c.m. energy range 1.05--2.00 GeV. The data with an integrated luminosity
of 34.6~pb$^{-1}$ collected with the SND detector at the VEPP-2000
$e^+e^-$ collider VEPP-2000 in 2011 and 2012 have been used for this 
measurement. The obtained cross section has a complex energy dependence
(see. Fig.~\ref{crsecb}) explained by the fact that all resonances of
the  $\rho$, $\omega$, and $\phi$ families give contributions to the
$e^+e^-\to K^+K^-$ amplitude.

\begin{figure}
\includegraphics[width=0.9\textwidth, trim=0mm 20mm 0mm 0mm]{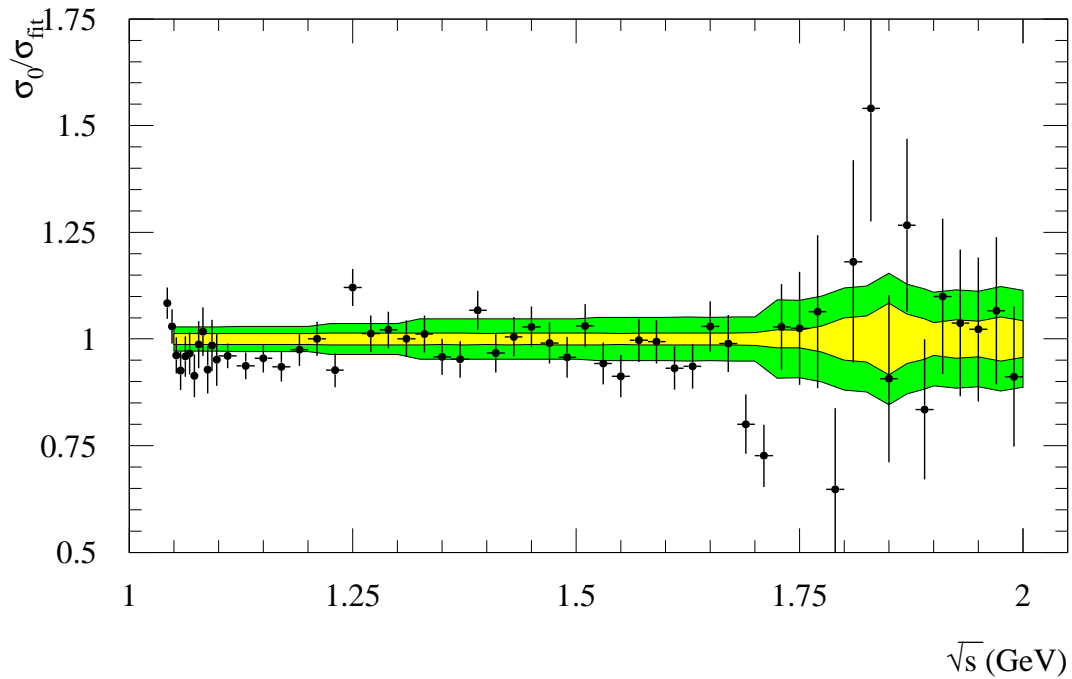}
\caption{\label{reldif}
The relative difference between the $e^+e^-\to K^+K^-$ cross section 
measured in the BABAR experiment~\cite{babar} and the fit to SND data
shown in Fig.~\ref{crsecb}. For BABAR data (points with error bars),
the statistical errors (diagonal elements of the covariance
matrix) are shown. The SND and BABAR systematic uncertainties are shown
by the light and dark shaded bands, respectively.
}
\end{figure}
The comparison of our measurement with the most accurate previous measurement
of the $e^+e^-\to K^+K^-$ cross section in the BABAR  experiment~\cite{babar}
is shown in Fig.~\ref{crsecb} and \ref{reldif}. Our measurement 
is in good agreement with BABAR data and has comparable or better accuracy.
We confirm the disagreement with the SND@VEPP-2M~\cite{sndkc} and
DM2~\cite{dm2} $e^+e^-\to K^+K^-$ data observed by BABAR.

\section{ACKNOWLEDGMENTS}
This work is supported in part by the RFBR grants 
16-02-00327-a, 16-02-00014-а and 15-02-01037.

\end{document}